\documentclass[prd,11pt,tightenlines,noshowpacs,nofootinbib,amsfonts,amsmath,amssymb,superscriptaddress,onecolumn,aps]{revtex4-1}

\usepackage{epsfig,amsmath,amssymb,verbatim,mathrsfs,hyperref}

\def\beq{\begin{eqnarray}}
\def\eeq{\end{eqnarray}}
\def\bea{\begin{eqnarray}}
\def\eea{\end{eqnarray}}

\def\gev{\, {\rm GeV}}

\newcommand{\gsim}{\lower.7ex\hbox{$\;\stackrel{\textstyle>}{\sim}\;$}}
\newcommand{\lsim}{\lower.7ex\hbox{$\;\stackrel{\textstyle<}{\sim}\;$}}

\def\mpl{M_{\rm Pl}}

\newcommand{\nnmb}{\nonumber}

\newcommand{\lrf}[2]{\left(\frac{#1}{#2}\right)}

\newcommand{\lag}{\mathscr{L}}

\newcommand{\bpsi}{\overline{\psi}}
\newcommand{\mpsi}{m_{\psi}}
\newcommand{\npsi}{n_{\psi}}
\newcommand{\npsieq}{n_{\psi,eq}}

\graphicspath{{./}{./Figs/}}

\begin{document}

\setlength{\baselineskip}{0.2in}



\title{Infrared Effects of Ultraviolet Operators\\on Dark Matter Freeze-In}
\author{Lindsay Forestell}
\affiliation{TRIUMF, 4004 Wesbrook Mall, Vancouver, BC V6T 2A3, Canada}
\affiliation{Department of Physics and Astronomy, University of British Columbia, Vancouver, BC V6T 1Z1, Canada}
\author{David E. Morrissey}
\affiliation{TRIUMF, 4004 Wesbrook Mall, Vancouver, BC V6T 2A3, Canada}
\date{\today}
\begin{abstract}
Dark matter~(DM) that interacts too weakly with the Standard Model~(SM)
to reach full thermodynamic equilibrium can be still be created in
significant amounts by rare SM collisions.  This mechanism, called freeze-in,
can proceed through a renormalizable connector operator with a very small
coefficient, or a non-renormalizable connector operator suppressed by a large
mass scale.  In the latter non-renormalizable scenario, 
the dominant creation of DM particles typically occurs at the largest
SM temperature attained during the radiation era 
(assuming a standard cosmological history), and for this
reason it is referred to as ultraviolet freeze-in.  We show that 
non-renormalizable operators can also contribute importantly
to the DM density at lower temperatures down to below the mass of the DM
particle.  To do so, we compute the production, annihilation, and freeze-out
of DM in a simple dark sector consisting of a massive Dirac fermion
DM candidate coupled to a massless Abelian vector boson with the only 
connection to the SM through the fermionic Higgs portal operator.  
For a broad range of parameters in the theory, the dark sector is populated
by ultraviolet freeze-in in the usual way, self-thermalizes to a dark
temperature below that of the SM, and undergoes thermal freeze-out.
We show that late residual freeze-in reactions during the freeze-out 
process can further populate the dark sector and increase the DM relic
density beyond standard dark sector freeze-out.
\end{abstract}
\maketitle

\section{Introduction\label{sec:intro}}

It has long been known that the Standard Model~(SM) does not provide a complete 
description of the universe. A key missing element is dark matter~(DM), 
which has been observed cosmologically to make up the majority 
of matter today~\cite{Aghanim:2018eyx}. 
However, very little is known about DM beyond its gravitational influence,
such as its particle properties or how its density was created 
in the early universe~\cite{Jungman:1995df,Bertone:2004pz,Lisanti:2016jxe}.  

Many theories of DM coupled directly to the SM 
rely on thermal production, with the most-studied paradigm being
thermal freeze-out~\cite{Wolfram:1978gp,Scherrer:1985zt,Kolb:1990vq}.  
In this process, the DM species begins 
in thermodynamic equilibrium with the rest of the SM at temperatures
above its mass.  As the universe cools below the DM mass, 
annihilation to SM particles allows the DM density
to track the exponentially suppressed equilibrium value until it becomes
too slow to keep up with the cosmological expansion.  
This simple mechanism for DM production has many attractive features:
it is insensitive to the state of the very early universe, 
and it yields the the correct relic abundance (to within a couple orders
of magnitude) for a generic weakly-interacting massive particle~(WIMP) 
with mass near the weak scale.

  Despite these features of thermal freeze-out, 
the lack of discovery in direct detection experiments and collider searches 
for WIMPs has motivated the study of other DM production 
mechanisms~\cite{Dev:2013yza,Baer:2014eja}.   
A promising alternative is freeze-in~(FI)~\cite{Hall:2009bx}, 
in which the DM species is assumed interact only very feebly with 
the SM and to have an initial abundance well below the value it would 
obtain in equilibrium with the SM plasma.  
Transfer reactions of the form SM+SM $\to$ DM+DM then create a 
sub-equilibrium abundance that evolves to the DM density seen today.

Within this paradigm, there are two general classes of connectors
between DM and the SM with very different cosmological behaviors. 
The first and most studied has DM connected to the SM through a 
very small renormalizable operator.  Production of DM for this class
is dominated by temperatures near the DM mass, $T \sim \mpsi$~\cite{McDonald:2001vt,Hall:2009bx,Cheung:2010gj,Cheung:2010gk,Chu:2011be,Yaguna:2011qn,Yaguna:2011ei,Blennow:2013jba,Heikinheimo:2016yds,Krnjaic:2017tio,Baker:2017zwx,Bernal:2017kxu}. 
For this reason, it is usually categorized as infrared~(IR) freeze-in.
This FI mechanism retains much of the attractive insensitivity to initial
conditions as WIMP freeze-out aside from the assumption of a very small 
initial DM density.  On the flip side, the renormalizable couplings 
needed for IR freeze-in must be extremely feeble.

The second class of connectors leading to freeze-in are non-renormalizable
operators connecting the DM to the SM, whose interaction strength is naturally
very small at low temperatures. Dominant DM production typically occurs 
at the highest SM temperatures attained during the radiation era,
and for this reason they lead to what is called ultraviolet~(UV) 
freeze-in~\cite{Elahi:2014fsa,Roland:2014vba,McDonald:2015ljz,Chen:2017kvz,Garny:2018grs}, 
with a well-known example being the 
gravitino~\cite{Pagels:1981ke,Ellis:1984er,Berezinsky:1991kf,Moroi:1993mb}.
A less attractive property of this paradigm, however, is that the DM
abundance depends on the state of the universe very early in its history.

  In this work we demonstrate that both UV and IR freeze-in can play a
role in determining the DM relic abundance through a single, non-renormalizable
connector operator.  This contrasts with the standard expectation that
non-renormalizable operators decouple once and for all at higher temperatures.
We illustrate this feature in a simple dark sector model consisting
of a stable Dirac fermion $\psi$ with mass $\mpsi$ that is charged
under an unbroken $U(1)_x$ gauge force with vector boson $X^{\mu}$
and coupling strength $\alpha_x = g_x^2/4\pi$. 
The only connection between the dark sector and the Standard Model~(SM)
is assumed to be through the \emph{fermionic Higgs portal} operator,
\beq
-\lag \supset \frac{1}{M}|H|^2\,\bpsi\psi \ .
\label{eq:fhp}
\eeq
Here, $M$ defines a very large mass scale of new physics above the energy
and temperature ranges we consider.  Note that we assume no
gauge kinetic mixing between $U(1)_x$ and hypercharge,
which can be enforced by an exact charge conjugation symmetry 
in the dark sector~\cite{DiFranzo:2015nli}.

  The UV connector operator of Eq.~\eqref{eq:fhp} can generate 
both UV and IR freeze-in effects over a broad range of parameters
when three plausible conditions are met.  First, reheating
after inflation is assumed to populate only the SM sector
with visible reheating temperature $T_{RH}$ well below the connector
mass scale $M$.  The dominant source of dark sector particles then comes
from visible-to-dark transfer reactions through the connector 
operator (UV freeze-in).
Second, for moderate to large values of the dark sector gauge coupling 
$\alpha_x$ the dark sector can self-thermalize to a temperature
$T_x$ less than the visible temperature $T$ but greater or similar
to the dark fermion mass $\mpsi$.  And third, if the DM annihilation
cross section is sufficiently large the DM abundance can track the
equilibrium abundance (at temperature $T_x < T$) for long enough
that transfer reactions from the non-renormalizable connector operator
return as the dominant contributor to the DM abundance.
To the best of our knowledge, combined UV and IR freeze-in effects
have not been investigated before, and they provide a counterexample
to the standard decoupling of non-renormalizable operators 
in the early universe.

  The combined UV and IR freeze-in behavior we focus on in the present
work is only one of a number of ``phases'' of freeze-out and freeze-in
possible within this simple dark sector model.  These phases are analogous
to the four phases studied in Ref.~\cite{Chu:2011be} for a similar
dark sector consisting of a charged complex scalar DM particle
connected to the SM Higgs field through the standard renormalizable
Higgs portal operator, but tilted towards the UV.  
When the mass scale $M$ in the fermionic connector operator of Eq.~\eqref{eq:fhp} 
is large relative to the weak scale and $\alpha_x \to 0$, 
the theory reduces to standard UV freeze-in of $\psi$ dark matter as studied 
in Ref.~\cite{McDonald:2015ljz} with no significant dark self-thermalization
or later annihilation.  In contrast, for much smaller $M$ near the TeV scale
the dark and visible sectors are thermally coupled (via the connector)
throughout $\psi$ freeze-out, and this operator can control
the freeze-out process even when $\alpha_x$ is 
very small~\cite{LopezHonorez:2012kv,deSimone:2014pda,Fedderke:2014wda}.
We focus on the scenario between these relative extremes with larger $M$
and $\alpha_x$.

This paper is structured as follows. Following the introduction,
we discuss in Sec.~\ref{sec:transfer} the UV freeze-in transfer of number and 
energy density through the connector operator of Eq.~\eqref{eq:fhp} 
as well as dark self-thermalization.  Next, in Sec.~\ref{sec:dm}
we compute the interplay between freeze-out and IR freeze-in in determining
the relic abundance $\psi$ particles and determine the conditions under
which both UV and IR freeze-in can be relevant.  
In Sec.~\ref{sec:dmself}, we comment briefly on the astrophysical implications 
of the new dark force from dark matter self-interactions.  
Finally, Sec.~\ref{sec:conc} is reserved for our conclusions.
Some technical details related to thermally-averaged cross sections
are contained in Appendix~\ref{sec:appa}.

\section{Populating the Dark Sector through UV Freeze-In\label{sec:transfer}}

  We begin by investigating the transfer of energy and number density to
the dark sector by UV freeze-in through the connector operator 
of Eq.~\eqref{eq:fhp}.  For this, we make the standard freeze-in assumption
that only the visible SM sector is populated significantly by reheating
after inflation with reheating temperature 
$T_{RH} \ll M$~\cite{Hall:2009bx,Elahi:2014fsa}.\footnote{Obtaining 
such an asymmetric reheating between different sectors has been studied
recently in Refs.~\cite{Adshead:2016xxj,Hardy:2017wkr}.}
The dark sector is then populated by transfer reactions of the 
form $H+H^{\dagger} \to \psi+\bpsi$ (assuming unbroken electroweak) 
mediated by the operator of Eq.~\eqref{eq:fhp}.
Once the number density of $\psi$ grows large enough, the dark sector
may also thermalize to an effective temperature $T_x$ 
through further reactions such as $\psi+\bpsi \leftrightarrow X^{\mu}+X^{\nu}$.
In this section we study the creation of $\psi$ particles from SM collisions
during and after reheating as well as the conditions for the self-thermalization
of the dark sector.

\subsection{Transfer without the Dark Vector}

  It is convenient to study first the creation of $\psi$ fermions
by SM collisions in the absence of dark vectors 
($\alpha_x\to 0$)~\cite{McDonald:2015ljz}.
The number and energy transfer via $H+H^{\dagger}\to \psi+\bpsi$ is described by
\beq
\frac{d\npsi}{dt} &=& -3H\npsi 
- \langle\sigma_{tr} v(T)\rangle(\npsi^2-\npsieq^2(T))\\
\frac{d\rho_x}{dt} &=& -3H\rho_x
- \langle\Delta E\cdot\sigma_{tr} v (T)\rangle(\npsi^2-\npsieq^2(T)) 
\eeq
where $\rho_x$ is the total energy density in the dark sector
and $\Delta E$ is the energy transfer per collision.

  Starting with number transfer, in the limit of $\npsi \ll \npsieq$
and $T \gg \mpsi$ the collision term is approximately
\beq
-\langle\sigma_{tr} v(T)\rangle(\npsi^2-\npsieq^2(T)) 
~\simeq~ 
\frac{1}{4\pi^5}\frac{T^6}{M^2} \ .
\eeq
Details of the calculation are given in Appendix~\ref{sec:appa}.
Assuming radiation domination up to the reheating temperature $T_{RH}\gg \mpsi$,
this gives the simple solution for the yield of $\psi$ (and $\bpsi$) of
\beq
Y_\psi(T) ~\simeq~ Y_\psi(T_{RH}) 
+ Y_{\psi,eq}(T)\,\frac{\sqrt{5/2}}{2\,\zeta(2)\,\pi^4}\;g_*^{-1/2}\,
\frac{\mpl T_{RH}}{M^2}
\left[1 - \lrf{T_{RH}}{T}^{-1}\right] \ .
\label{eq:ypsirh}
\eeq
This solution only holds in the limit $Y_{\psi} \ll Y_{\psi,eq}$,
corresponding to a consistency condition of
(for $Y_{\psi}(T_{RH}) \to 0$ and $T \ll T_{RH}$)\footnote{The number
and energy density produced through thermal transfer prior to reheating
by the operator of Eq.~\eqref{eq:fhp}
is a very small fraction of that produced 
at reheating~\cite{McDonald:2015ljz,Chen:2017kvz}.}
\beq
T_{RH} \ll \frac{2\,\zeta(2)\pi^4}{\sqrt{5/2}}\;g_*^{1/2}\,\frac{M^2}{\mpl} \ .
\eeq
Larger reheating temperatures imply thermalization between the dark and
visible sectors at reheating with $Y_{\psi}(T) \to Y_{\psi,eq}(T)$
for $T\sim T_{RH}$.  In this work we focus on the non-thermalization scenario.

Turning next to energy transfer, the transfer term is 
computed in Appendix~\ref{sec:appa} and for $\mpsi \ll T \ll M$
and $\npsi\ll\npsieq$ reduces to
\beq
-\langle\Delta E\cdot\sigma_{tr} v (T)\rangle(\npsi^2-\npsieq^2(T)) 
~\simeq~
\frac{3}{2\pi^5}\frac{T^7}{M^2} \ .
\eeq
Solving as above, we find
\beq
\lrf{\rho_x}{\rho_{\psi,eq}} ~\simeq~
\lrf{\rho_x}{\rho_{\psi,eq}}_{T_{RH}} +
\frac{180\sqrt{10}}{7\pi^8}\,g_*^{-1/2}
\frac{\mpl T_{RH}}{M^2}
\left[1 - \lrf{T_{RH}}{T}^{-1}\right] \ .
\label{eq:rhoxrh}
\eeq
Again, this is only valid for $Y_{\psi} \ll Y_{\psi,eq}$.  
For sufficiently large $T_{RH}$, $\rho_x \to \rho_{\psi,eq}(T)$ at
$T\sim T_{RH}$.

  Comparing $Y_{\psi}$ and $\rho_x$ found above for $Y_\psi \ll Y_{\psi,eq}$,
we see that the mean momentum of the fermions produced near reheating is on the
order $p \sim T_{RH}$.  At later times, these momenta simply redshift 
as $1/a$ provided $T \gg \mpsi$.  Indeed, the detailed analysis 
of Ref.~\cite{McDonald:2015ljz} shows that (in the absence of dark vectors) 
the dark fermions obtain an approximate \emph{Bose-Einstein} distribution 
with effective temperature $T_x \simeq (1.155)\,T_{RH}(a_{RH}/a)$.

\subsection{Thermalization with the Dark Vector}

  Let us now include a dark vector boson $X^\mu$ coupling to $\psi$ with strength
$\alpha_x=g_x^2/4\pi$.  This interaction allows the dark fermions to scatter 
with each other, annihilate to vector bosons, and emit vectors 
as radiation.  If these reactions are strong enough, the dark fermion 
and vector species can thermalize with each other to yield an 
effective temperature $T_x \leq T$.  

The self-thermalization of heavy dark particles coupled to a massless dark vector
was investigated in Refs.~\cite{Chu:2011be,Garny:2018grs}.  As in these works,
we only make parametric estimates of the very complicated full thermalization
processes.  We identify self-thermalization in the dark sector
with the condition
\beq
\Gamma_{th}(T_{th}) = H(T_{th}) \ ,
\eeq
where $\Gamma_{th}$ is an effective thermalization rate to be discussed
below and this relation defines the visible thermalization 
temperature $T_{th}$ implicitly.  Note that $T_{th} \leq T_{RH}$,
and we set $T_{th} = T_{RH}$ if $\Gamma_{th}(T_{RH}) \geq H(T_{RH})$.

  It is convenient to classify the thermalization processes contributing
to $\Gamma_{th}$ into: i) $2\to 2$ processes with hard momentum exchange; 
ii) $2\to 3$ inelastic processes together with $2\to 2$ with soft momentum
exchange.  The first class includes annihilation $\psi+\bpsi \to X^\mu+X^\nu$
and hard scatterings such as $\psi+\psi\to \psi+\psi$ for which we estimate the
rate to be~\cite{Chu:2011be}
\beq
\Gamma_{el}(T) ~\sim~ \frac{\pi\,\alpha_x^2}{T^2}\;n_{\psi}(T) \ ,
\label{eq:gamel}
\eeq
where $\npsi(T)$ is the number density of $\psi$ prior to dark 
self-thermalization.  Using Eq.~\eqref{eq:ypsirh} 
(with $Y_{\psi}(T_{RH})\to 0$), for $T\gg \mpsi$ it is given by
\beq
n_{\psi}(T) ~\simeq~ \frac{3\sqrt{5/2}}{2\,\pi^6g_*^{1/2}}\,
\frac{\mpl T_{RH}}{M^2}\;T^3 \ .
\eeq
The second class of soft and inelastic processes was studied 
in Ref.~\cite{Garny:2018grs} with the net result
\beq
\Gamma_{in}(T) ~\sim~ \min\Bigg\{\frac{\alpha_x^3\npsi(T)}{\mu_{IR}^2},\,
\alpha_x^2\sqrt{\npsi/T}\Bigg\} \ ,
\label{eq:gamin}
\eeq
where $\mu_{IR}$ an effective infrared cutoff given by
\beq
\mu_{IR} = \max\Bigg\{\sqrt{\alpha_x\npsi/T},\, H,\,m_\psi\Bigg\} \ .
\eeq
We take the full thermalization rate to be the sum of 
the hard and inelastic rates, $\Gamma_{th} = \Gamma_{in}+\Gamma_{el}$.

  If thermalization occurs with $T_{th} \gg \mpsi$, a smaller number
of $\psi$ and $\bpsi$ fermions with typical energy $T$ are redistributed
into a larger number of $\psi$, $\psi$, and $X^{\mu}$ particles
in equilibrium with each other at temperature $T_x$.  Treating the thermalization
as instantaneous, the resulting dark sector temperature can be obtained
from energy conservation and the result of Eq.~\eqref{eq:rhoxrh}:
\beq
\frac{T_x(T_{th})}{T_{th}} ~\equiv~ \xi(T_{th}) 
~\simeq~ \left[
\frac{180\sqrt{10}}{11\pi^8g_*^{1/2}}\,\frac{\mpl T_{RH}}{M^2}
\right]^{1/4}
 \ .
\label{eq:xith}
\eeq
At later times, separate conservation of entropy in the dark and visible sectors
implies
\beq
\xi(T) ~\simeq~ \xi(T_{th})\left[\frac{g_{*S}(T)}{g_{*S}(T_{th})}\cdot
\frac{g_{*S,x}(T_{th})}{g_{*S,x}(T)}
\right]^{1/3}
\ ,
\label{eq:xitt}
\eeq
where $g_{*S(x)}$ refers to the number of visible (hidden) 
entropy degrees of freedom.

  The analysis leading to the temperature ratio of Eq.~\eqref{eq:xith}
has three assumptions built into it, and their consistency
implies maximal and minimal allowed values of $\xi(T_{th})$.  
First, the assumption of non-thermalization 
between the visible and dark sectors implies $\xi(T_{th}) \ll 1$.  
Second, the validity of the effective connector operator description 
of Eq.~\eqref{eq:fhp} requires $T_{RH} \ll M$ corresponding to a maximum value
of $\xi(T_{th}) \lesssim (10^{-3}\,\mpl/M)^{1/4}$.  And third, we have so far
neglected the mass of the $\psi$ fermion.  Demanding that 
$T_x(T_{th}) \gtrsim \mpsi$ then leads to a lower bound on $\xi(T_{th})$
that we use to define
\beq
\xi_{min} \equiv \frac{\mpsi}{T_{th}} \ .
\eeq
This also defines an implicit lower bound on the thermalization temperature 
for given values of $\mpsi$, $M$, and $\alpha_x$, and correspondingly a lower
limit on the reheating temperature $T_{RH}$.

\begin{figure}[ttt]
	\centering
	\includegraphics[width=0.39\textwidth]{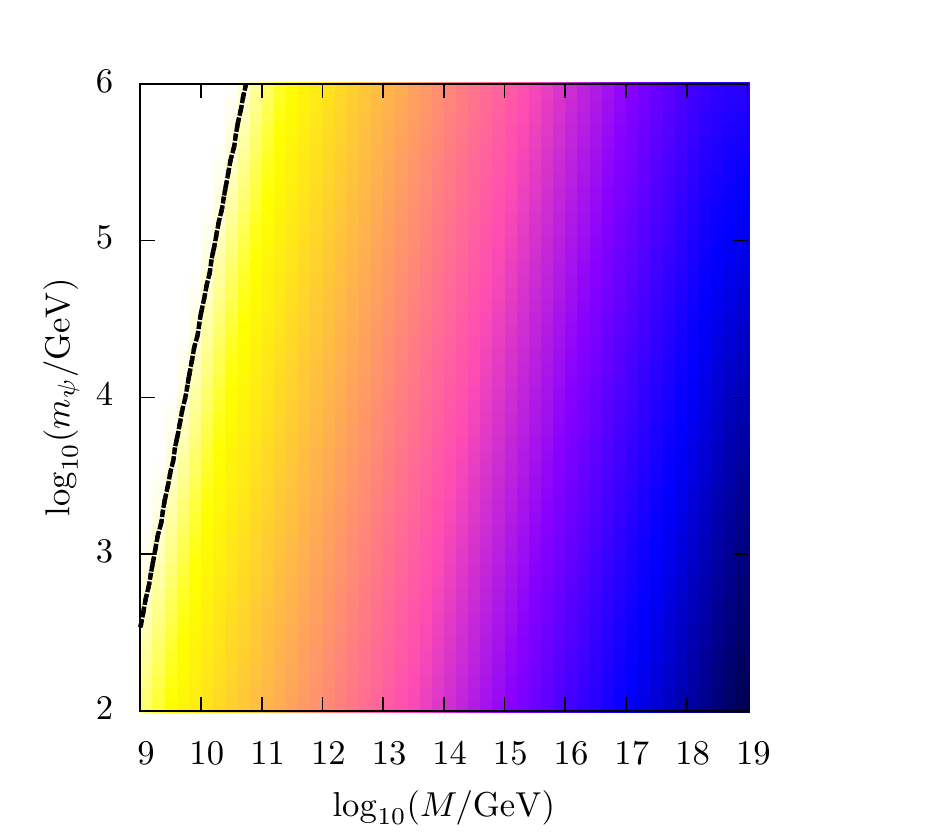}
        \hspace{-1.7cm}
	\includegraphics[width=0.39\textwidth]{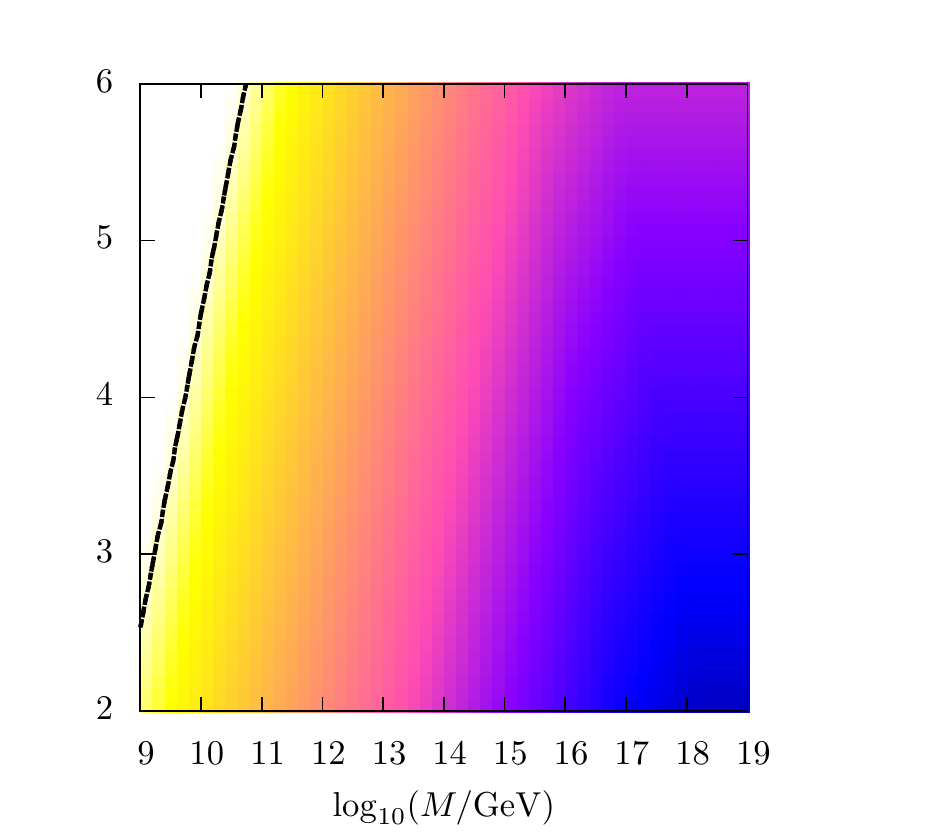}
        \hspace{-1.7cm}
	\includegraphics[width=0.39\textwidth]{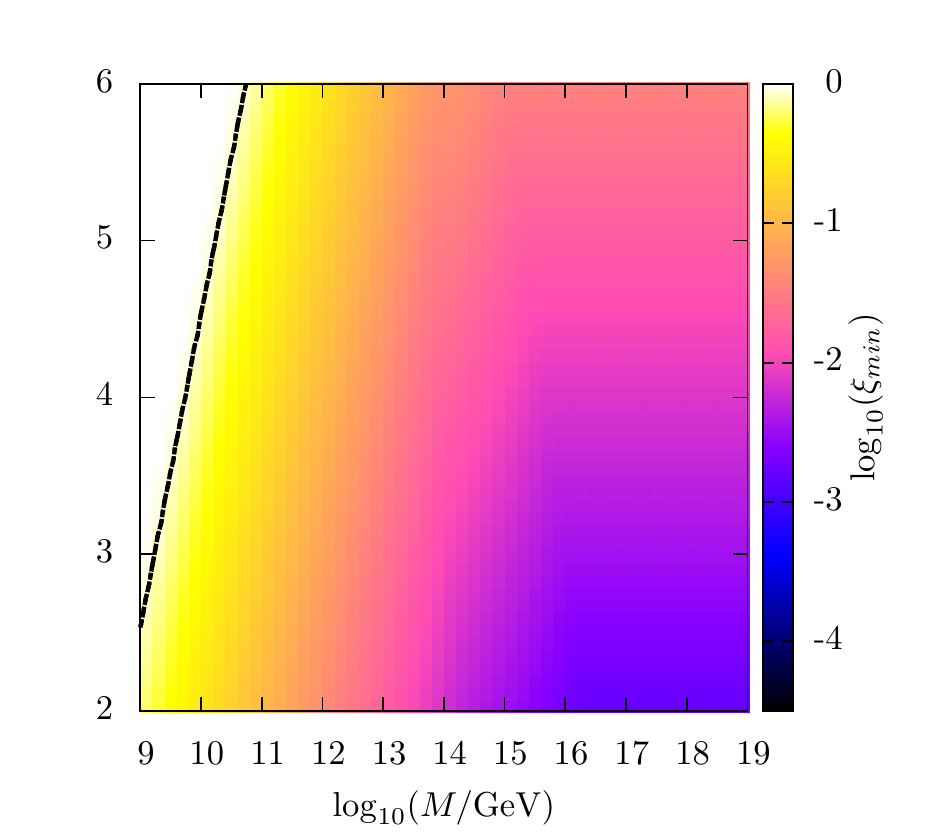}
	\caption{Minimum consistent values of $\xi(T_{RH})$ in the
$M$--$\mpsi$ plane for $\alpha_x=10^{-1}$~(left), 
$10^{-2}$~(middle), $10^{-3}$~(right).  The black line indicates where
$\xi(T_{RH}) \to 1$ and our assumption of non-thermalization with 
the SM breaks down.}
	\label{fig:cmax_regions}
\end{figure}

 In Fig.~\ref{fig:cmax_regions} we show the values of $\xi_{min}$ 
in the $M\!-\!\mpsi$ plane for $\alpha_x = 10^{-1}$~(left),
$10^{-2}$~(middle), and $10^{-3}$~(right).  Larger $M$ and $\alpha_x$
and smaller $\mpsi$ lead to smaller $\xi_{min}$.  
The white regions in the upper left corners of the plots
(bounded by black lines) have $\xi_{min}\to 1$ corresponding 
to thermalization between the visible and dark sectors when 
dark self-thermalization is achieved.  As stated above, in the analysis 
to follow we focus on the lower right region where this does not occur.

\section{Freeze-Out and Late Transfer in the Dark Sector\label{sec:dm}}

  If the dark sector is populated by UV freeze-in and is self-thermalized
at temperature $T_x \gtrsim \mpsi$, the dark fermion will undergo
freeze-out by annihilation to dark vectors when $T_x$ falls below $\mpsi$.  
While freeze-out in a dark sector with $T_x \ll T$ has been studied
in Refs.~\cite{Feng:2008mu,Das:2010ts,Agrawal:2016quu,Garny:2018grs}, 
we point out a qualitatively new feature in
the present context.  Specifically, we show that the UV connector operator
responsible for initially populating the dark sector at reheating 
can drastically change the freeze-out dynamics at much later times.

\subsection{Evolution Equations}

  The evolution of the $\psi$ dark fermion number density at $T_x \lesssim \mpsi$
is described by
\beq
\frac{d{n}_{\psi}}{dt}  +3H\,\npsi ~\simeq~   
-\langle\sigma v(T_x)\rangle_{ann}(\npsi^2-n^2_{\psi,eq}(T_x))
+ \langle\sigma_{tr} v(T)\rangle n^2_{\psi,eq}(T)
\label{eq:nann}
\eeq
In writing this expression we have assumed self-thermalization in the dark
sector with $T_x \ll T$ and no asymmetry between $\psi$ and $\bpsi$.

The first term on the right side of Eq.~\eqref{eq:nann} describes
annihilation $\psi+\bpsi\to X^\mu+X^\nu$ with a thermal average at 
temperature $T_x$.  The leading-order perturbative result for the
cross section at low velocity is~\cite{Pospelov:2007mp}
\beq
\sigma_{ann,p}v = \frac{\pi\,\alpha_x^2}{\mpsi^2} \ .
\eeq
However, the full cross section receives independent non-perturbative enhancements
from the Sommerfeld effect~\cite{Hisano:2003ec,Hisano:2004ds,Cirelli:2007xd} 
and bound state formation~\cite{Feng:2008mu,Pospelov:2008jd}.
The full cross section can be written in the 
form~\cite{Feng:2008mu,vonHarling:2014kha}
\beq
\sigma_{ann}v = 
\left[\mathcal{S}_{somm}(v)+\mathcal{S}_{rec}(v)\right]\,\sigma_{ann,p}v \ ,
\eeq
where $v$ is the relative velocity and
\beq
\mathcal{S}_{som}(v) &=& \frac{2\pi z}{1-e^{-2\pi z}} \ , \\
\mathcal{S}_{rec}(v) &=& \mathcal{S}_{som}(v)\,\frac{2^9}{3~}\,
\frac{z^4}{(1+z^2)^2}\,
\,e^{-4z\tan^{-1}(1/z)}
 \ ,
\eeq
with $z = \alpha_x/v$, and which have the limits $\mathcal{S}_i(v) \to 1$ 
for $v \gg \alpha_x$.

The second term on the right side of Eq.~\eqref{eq:nann} corresponds to
transfer reactions of the form $H+H^{\dagger}\to \psi+\bpsi$, and has all relevant
quantities evaluated at the visible temperature $T$.\footnote{Since $T_x\ll T$,
we can neglect the reverse reaction.}
An explicit expression for this transfer term is given in Appendix~\ref{sec:appa},
which reduces to
\beq
\langle\sigma_{tr} v(T)\rangle n^2_{\psi,eq}(T) 
~\equiv~ \mathcal{T}(T)
~\simeq~
\left\{
\begin{array}{lcc}
\frac{1}{4\pi^5}\frac{T^6}{M^2}&;&T \ll \mpsi\\
&&\\
\frac{3}{32\pi^4}\frac{\mpsi^2T^4}{M^2}e^{-2\mpsi/T}&;&T \ll \mpsi
\end{array}\right.
\ . 
\label{eq:trop}
\eeq
For $T_x < \mpsi$ but $T\gg \mpsi$, the annihilation term in Eq.~\eqref{eq:nann}
receives an exponential suppression in temperature while the transfer term 
is only suppressed by a power.  We show below that this can allow the transfer
term derived from a UV connector operator to play a significant role in the IR.

\subsection{Analytic Estimates}

  It is instructive to estimate the relic density of $\psi$ particles
analytically to understand the effect of late-time transfer by the UV 
connector. To do so, we treat the annihilation cross section as being 
power-law in velocity: $\langle\sigma_{ann}v\rangle \to \sigma_0\,x_x^{-n}$
where $x \equiv \mpsi/T$ and $x_x \equiv \mpsi/T_x = \xi^{-1}\,x$.

{\begin{flushleft}
\textbf{Freeze-Out Without the Transfer Term}
\end{flushleft}\vspace{-0.35cm}
  Consider first the relic density of $\psi$ with no transfer term 
but a definite value of $\xi \ll 1$.  This can be computed by a 
simple generalization~\cite{Feng:2008mu,Das:2010ts,Agrawal:2016quu} 
of the analytic freeze-out approximation 
of Refs.~\cite{Wolfram:1978gp,Scherrer:1985zt,Kolb:1990vq,Gondolo:1990dk,Edsjo:1997bg}.
Freeze-out occurs when the mass to dark temperature ratio is
\beq
{x}_{x,fo} ~\simeq~ 
\ln\left[(0.192)(n+1)(g_\psi/g_*^{1/2})\mpl\,\mpsi\,\sigma_0\,\xi^2\right]
-(n+\frac{1}{2})\ln({x}_{x,fo})
\ ,
\label{eq:xxfo}
\eeq
which can be solved iteratively for $x_x^{fo}$.  
This translates into an approximate relic density of\footnote{
Note that we use $\mpl = 2.43\times 10^{18}\,\gev$,
and the full DM relic density is the sum of equal $\psi$ and $\bpsi$ densities.}
\beq
\Omega_{\psi}h^2 ~\simeq~
(2.07\times 10^{8}\,\gev^{-1})\,
\frac{\xi\,(n+1)\,{x}^{n+1}_{x,fo}}{(g_{*S}/g_*^{1/2})\mpl\,\sigma_0}
\ .
\label{eq:omx1}
\eeq
Relative to the freeze-out of a species in thermodynamic equilibrium
with the visible sector with the same mass and cross section, 
these relations imply
\beq
x_x^{fo} ~\simeq~ \tilde{x}^{fo} + (2-1/\tilde{x}_{fo})\ln\xi \ ,~~~~~
\Omega_{\psi}h^2 ~\simeq~ 
\xi\left(1+2\ln\xi/\tilde{x}^{fo}\right)\widetilde{\Omega}_\psi h^2 \ ,
\eeq
where $\tilde{x}^{fo}$ and $\widetilde{\Omega}_\psi h^2$ are the values
for these quantities if the species were thermally coupled to the SM.
The most important change is a reduction of the relic
density by a factor of about $\xi \ll 1$.}

\begin{flushleft}
\textbf{Freeze-Out With the Transfer Term}
\end{flushleft}\vspace{-0.35cm}
Let us now include the transfer term from Eq.~\eqref{eq:nann}
in the evolution of the density of $\psi$.  As $T_x$ falls below
$\mpsi$, annihilation is expected to dominate and keep $\npsi$ close to 
its equilibrium value at temperature $T_x$.  However, 
since the corresponding annihilation rate falls exponentially in this regime, 
it decreases more quickly than the Hubble and transfer rates, 
and thus the near-equilibrium regime ends when one of these other rates 
catches up.  We show here that late-time transfer reactions can
significantly modify the final $\psi$ relic density when the annihilation
rate meets the transfer rate before reaching Hubble.

  Define $T_{x,=}$ to be the value of the dark temperature $T_x$
that solves the equation
\beq
\langle\sigma_{ann}v(T_x)\rangle\npsieq^2(T_x) 
= \mathcal{T}(T_x/\xi) \ ,
\eeq
where $\mathcal{T}(T)$ is the transfer rate given in Eq.~\eqref{eq:trop}.
If the solution has $T_= = T_{x,=}/\xi < \mpsi$, 
an approximate expression for it is
\beq
x_{x,=} ~\simeq~ \frac{1}{2}\ln
\left(\frac{\pi^2}{2}\,g_\psi^2\,\sigma_0M^2\,\xi^6\right)
+ \lrf{3-n}{2}\ln(x_{x,=})
\ ,
\eeq
which can be solved iteratively for $x_{x,=}$ provided it is greater than unity.
When $x_{x,=}$ is greater than the freeze-out temperature without transfer,
$x_{x,fo}$ given in Eq.~\eqref{eq:xxfo}, the transfer operator does
not significantly alter the $\psi$ relic density.  In particular, the condition
$x_{x,=} > x_{x,fo}$ implies that the evolution of the $\psi$ density
is dominated by Hubble dilution rather than transfer for all $x_x > x_{x,fo}$
since the expansion term decreases less quickly than the transfer term
in this regime.  In contrast, transfer effects are important
for $x_{x,=} < x_{x,fo}$.

  When $x_{x,=} < x_{x,fo}$, the transfer and annihilation terms 
in Eq.~\eqref{eq:nann} can reach a balance with each other 
for $x_x > x_{x,=}$ until the Hubble term catches up.  
The number density of $\psi$ is then approximately
\beq
n_{\psi,=}(T_x) &~\simeq~& \sqrt{\frac{\mathcal{T}(T_x/\xi)}{\sigma_0}}\;x_x^{n/2}
\label{eq:nbal}\\
&~\to~& \frac{1}{2\pi^{5/2}}\,
\frac{\mpsi^3}{\sqrt{\,\sigma_0M^2}}\;\xi^{-3}\,x_x^{-3+n/2}
\hspace{1cm}\left(T_x/\xi \gg \mpsi\right)
\eeq
where the expression in the second line only applies for $T_x/\xi \gg \mpsi$.
Note that the density in this regime is always greater than 
the equilibrium density $\npsieq(T_x)$, even when $T_x/\xi < \mpsi$.  

  If the balance regime is achieved, $x_{x,=} < x_{x,fo}$,
it ends when the Hubble term in Eq.~\eqref{eq:nann} catches up 
to the annihilation and transfer terms.
This later decoupling corresponds approximately to the condition
\beq
\langle\sigma_{ann}v(T_x)\rangle n_{\psi,=}(T_x) ~\simeq~ H(T_x/\xi) \ .
\label{eq:balfo}
\eeq
Defining $T_{x,dec}$ as the dark temperature that satisfies the relation
above, an approximate solution for $T_{x,dec}/\xi \gg \mpsi$ is
\beq
x_{x,{dec}} ~\simeq~ \left[(0.086)\,
\frac{\mpsi\mpl\sqrt{\sigma_0}}{g_*^{1/2}\,M}\;
\xi^{-1}\right]^{1/(1+n/2)} \ .
\label{eq:xxdec}
\eeq
The solution for $T_{x,dec}/\xi \lesssim \mpsi$ is more complicated but
can be obtained similarly.  The final relic density can be written
in a form very similar to standard freezeout via Eq.~\eqref{eq:balfo}:
\beq
\Omega_{\psi}h^2 ~\simeq~ 
(2.07\times 10^{8}\,\gev^{-1})\,
\frac{\xi\,{x}^{n+1}_{x,dec}}{(g_{*S}/g_*^{1/2})\mpl\,\sigma_0}
\ .
\label{eq:omx2}
\eeq
Since $n_{\psi,=}(T_x) > n_{\psi,eq}(T_x)$ we must have $x_{x,dec} > x_{x,fo}$
whether or not $T_{x,dec}/\xi$ is larger or smaller than $\mpsi$,
and therefore the relic density of Eq.~\eqref{eq:omx2} is bigger than
the pure freeze-out result of Eq.~\eqref{eq:omx1}.

\subsection{Numerical Results for Freeze-Out}

\begin{figure}[ttt]
	\centering
        \includegraphics[width=0.35\textwidth]{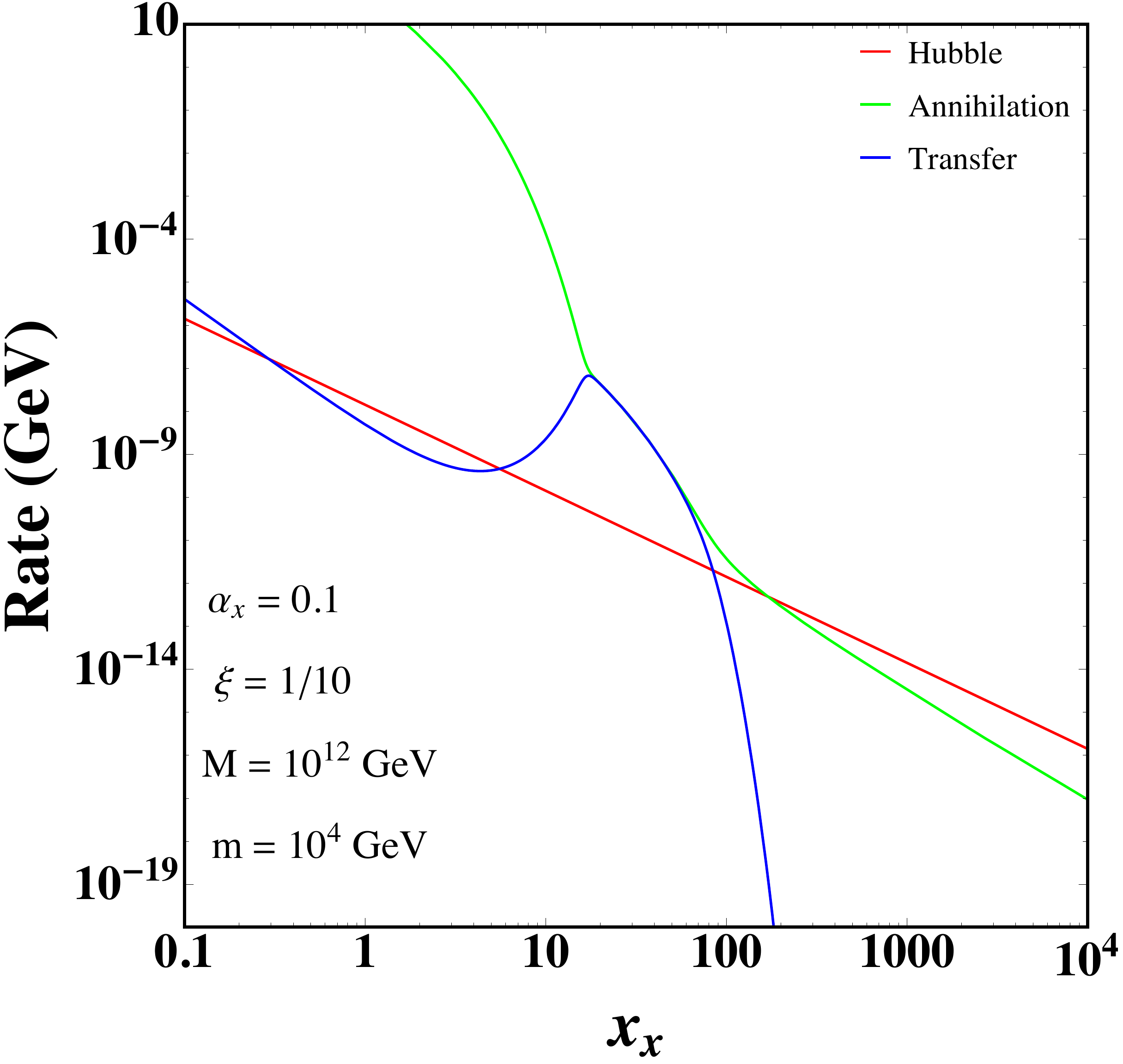}
        \hspace{1cm}
	\includegraphics[width=0.35\textwidth]{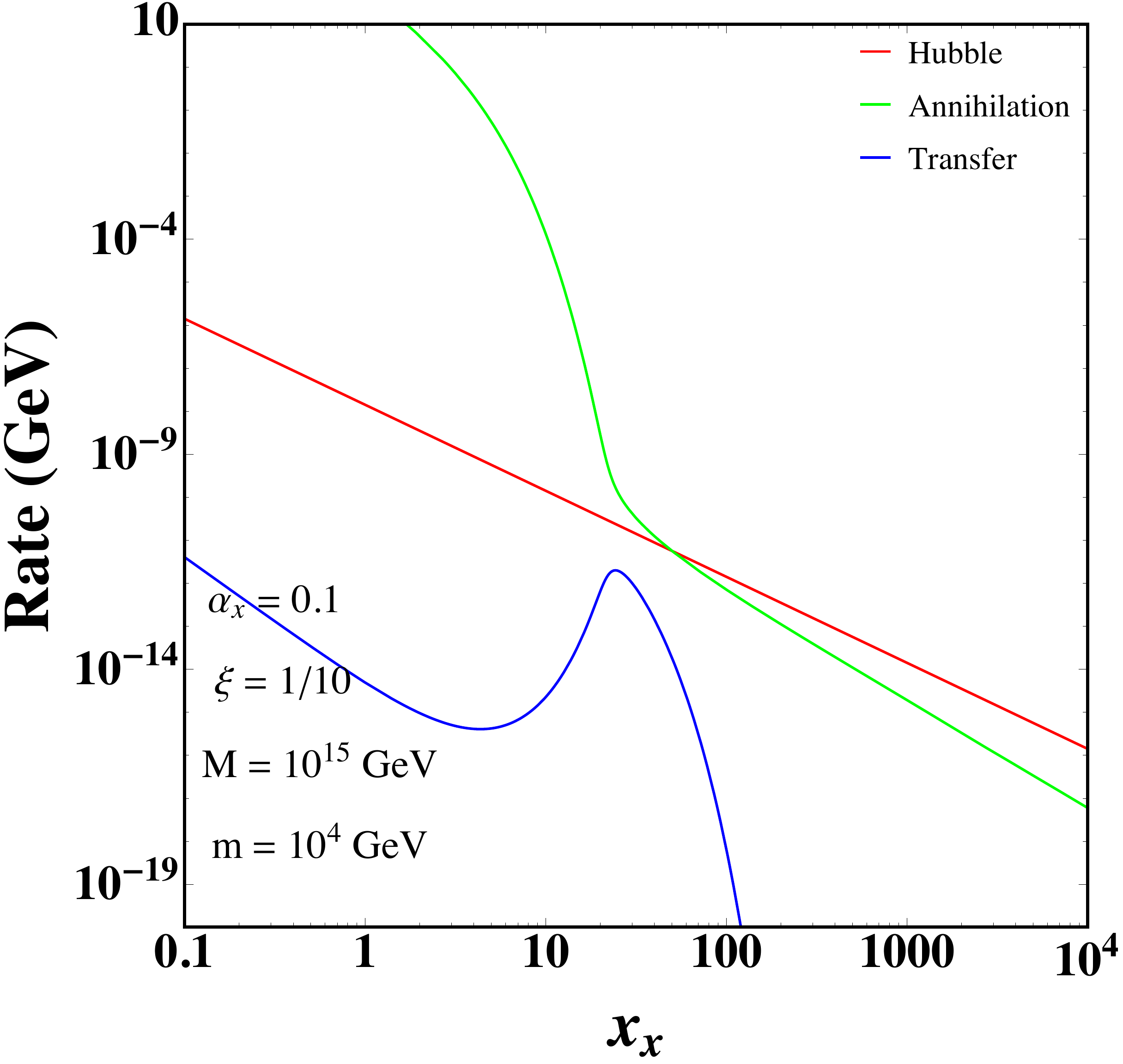}
        \vspace{0.5cm}\\
	\includegraphics[width=0.35\textwidth]{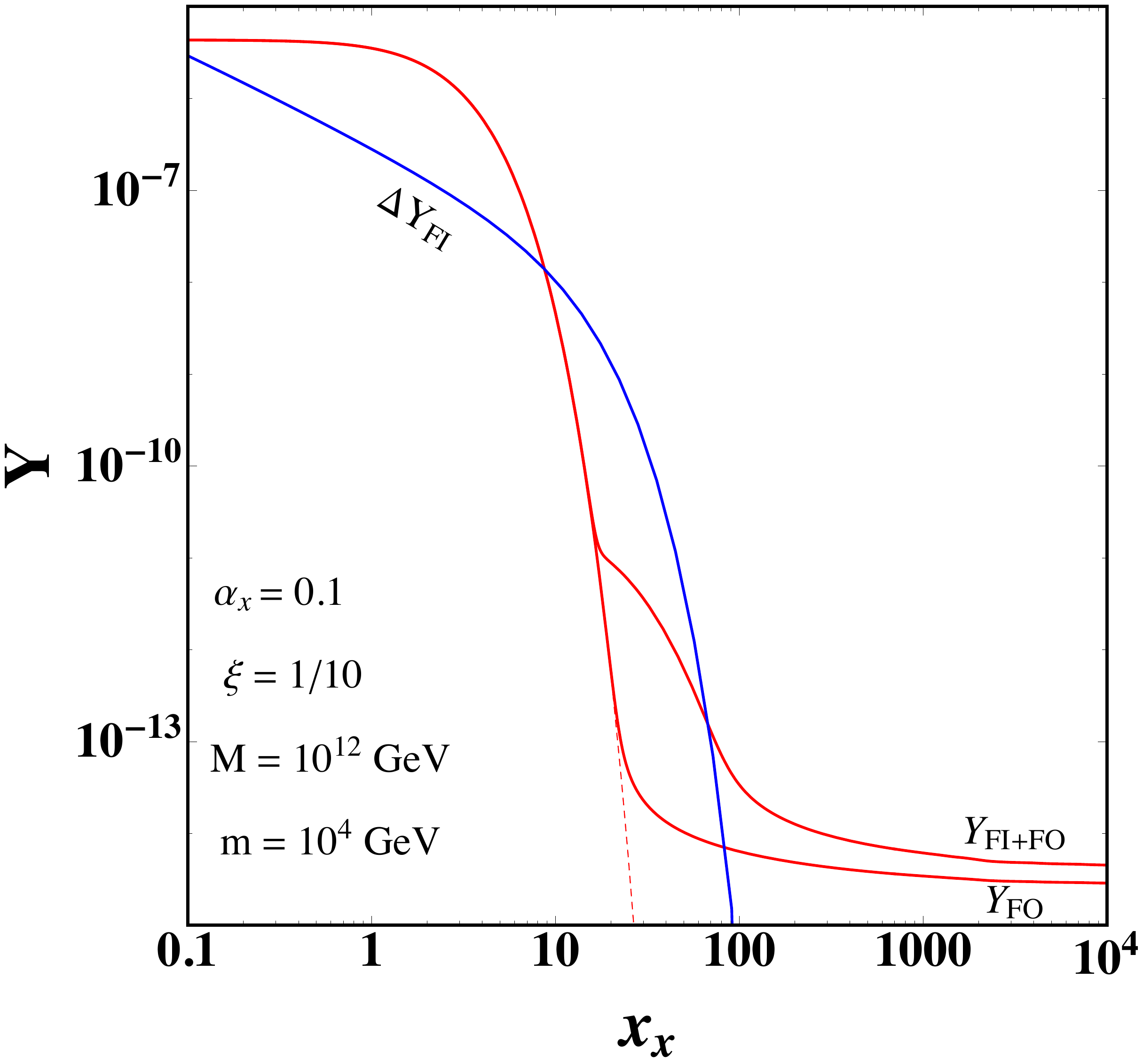}
        \hspace{1cm}
	\includegraphics[width=0.35\textwidth]{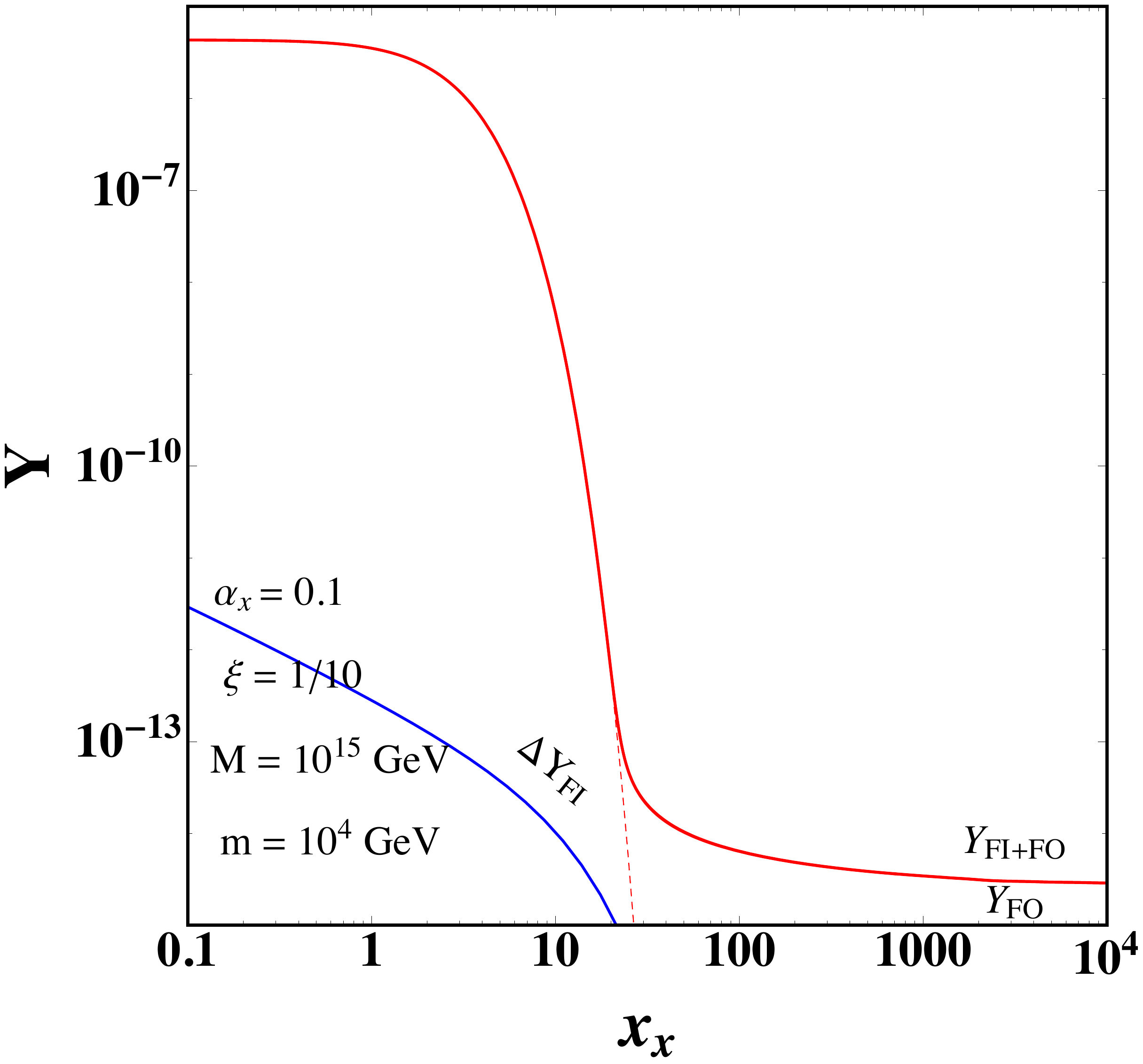}
	\caption{
Evolution of the relevant rates in the upper panels and the $\psi$ density in
the lower panels for $\alpha_x = 0.1$, $\xi = 0.1$, $\mpsi = 10^4\,\gev$, 
and $M = 10^{12}\,\gev$~(left) and $10^{15}\,\gev$~(right).
}
	\label{fig:fo_plots}
\end{figure}

  To confirm the analytic estimates derived above and map out the parameter
space of theory, we perform a full numerical analysis of the dark matter 
freeze-out process.  In Fig.~\ref{fig:fo_plots} we show the evolution
of the relevant rates in the upper panels and the $\psi$ density in
the lower panels for $\alpha_x = 0.1$, $\xi = 0.1$, $\mpsi = 10^4\,\gev$, 
and $M = 10^{12}\,\gev$~(left) and $10^{15}\,\gev$~(right).  
The rate plots show the rates for Hubble, annihilation, and late transfer
defined according to
\beq
\mathrm{Hubble} = H(T) \ ,~~~~~
\mathrm{Annihilation} = \langle\sigma_{ann} v(T_x)\rangle\,\npsi \ ,~~~~~
\mathrm{Transfer} = \mathcal{T}(T)/\npsi \ ,
\eeq
where $\npsi$ is the number density obtained from solving Eq.~\eqref{eq:nann}
and $\mathcal{T}(T)$ is the transfer rate of Eq.~\eqref{eq:trop}.
The value of $M$ is smaller in the left panels of this figure,
and late-time transfer 
In the $\psi$ number density plots, we show the densities 
in equilibirum~(dashed line), and with and without 
the transfer operator~(upper and lower solid lines).

  Late transfer by the fermionic Higgs portal operator is seen to increase
significantly the final relic density in the left panels 
of Fig.~\ref{fig:fo_plots}, while its effect is negligible in the right panels. 
The difference corresponds to the larger transfer rate for $M = 10^{12}\,\gev$ 
in the left panels versus $M=10^{15}\,\gev$ in the right.  
Following the rates for $M=10^{12}\,\gev$,
transfer is seen to catch up to annihilation before Hubble leading to a regime
of balanced rates and enhanced number density.  In contrast, the Hubble rate 
catches up to annihilation before transfer in the right panels with 
$M=10^{15}\,\gev$ and never plays a significant role in the evolution of $\npsi$.

\begin{figure}[ttt]
	\centering
	\includegraphics[width=0.39\textwidth]{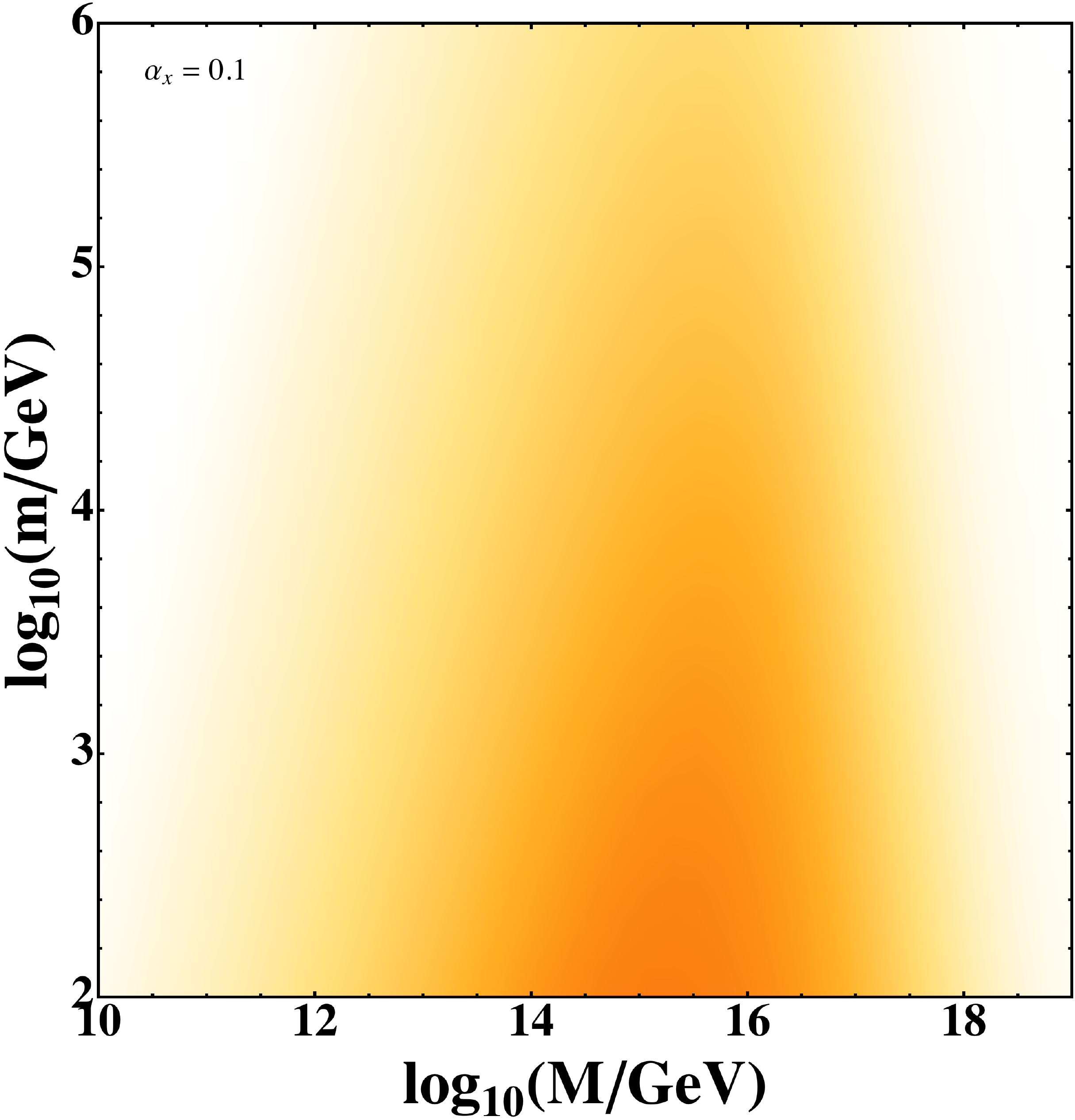}
	\hspace{0.5cm}
	\includegraphics[width=0.39\textwidth]{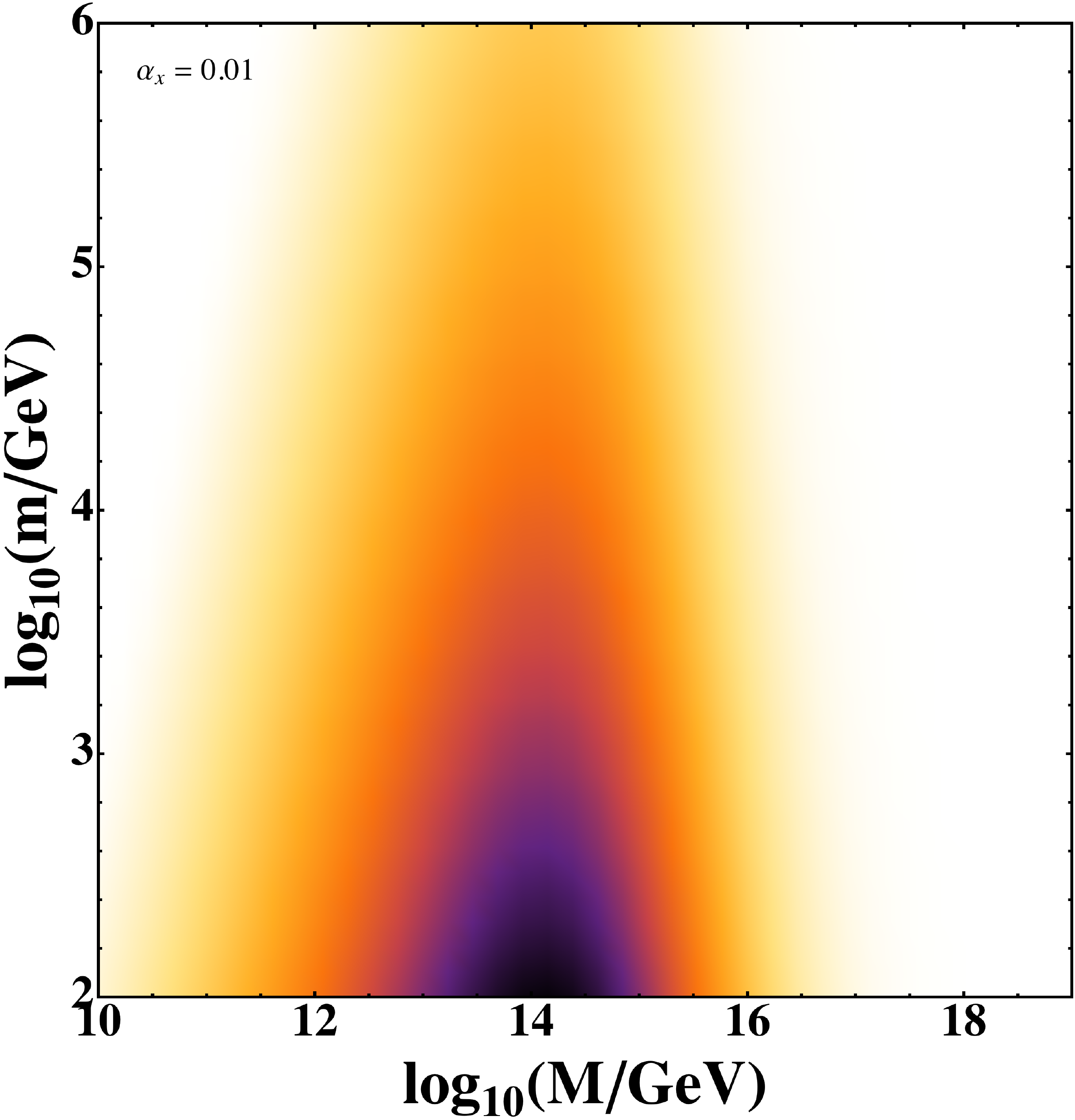}
        \hspace{0.3cm}
        \includegraphics[width=0.065\textwidth]{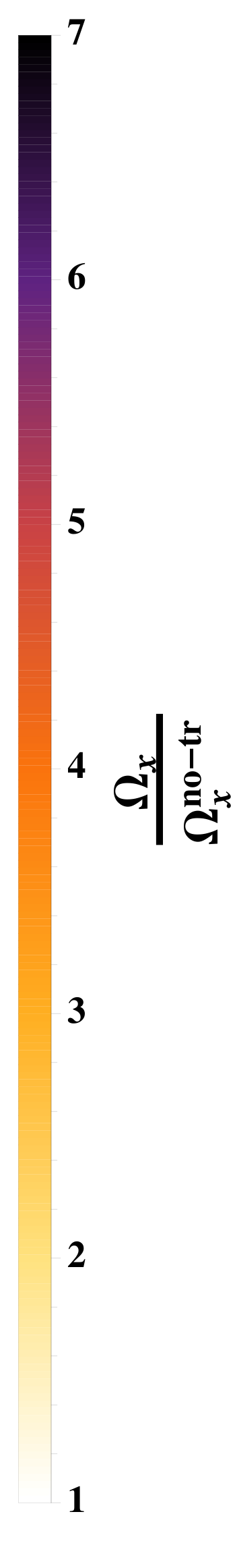}
	\caption{Enhancement of the $\psi$ relic density due to late transfer 
          effects relative to the value without this effect, 
          $\Omega_\psi/\Omega_\psi^{no-tr}$ for 
          $\alpha_x = 0.1$~(left) and $0.01$~(right) and $\xi = \xi_{min}$.
}
	\label{fig:heatmap_ratio}
\end{figure}

  In Fig.~\ref{fig:heatmap_ratio} we show the enhancement of the relic
density in the $M$--$\mpsi$ plane for $\alpha_x = 0.1$~(left) and
$0.01$~(right) with $\xi = \xi_{min}$ as computed previously.
The contours in both panels indicate the relic density we find to the value
that would be obtained without late transfer effects, 
$\Omega_\psi/\Omega_\psi^{no-tr}$.  Late transfer by the connector operator
initially increases as $M$ decreases and the transfer operators becomes
more effective. However, as $M$ continues to decrease
we find a competing effect between the efficiency of transfer 
and the increasing value of $\xi_{min}$.  
As the dark and visible temperatures approach
each other, transfer is more likely to occur while $T\to \mpsi$ and the
effect becomes exponentially suppressed, as seen in Eq.~\eqref{eq:trop}. 
Transfer effects are also reduced at $\alpha_x=0.1$ relative to $\alpha_x=0.01$
due to the non-pertubative enhancements in the annihilation cross section
at low velocities for the larger value of the gauge coupling.

\begin{figure}[ttt]
	\centering
        \includegraphics[width=0.45\textwidth]{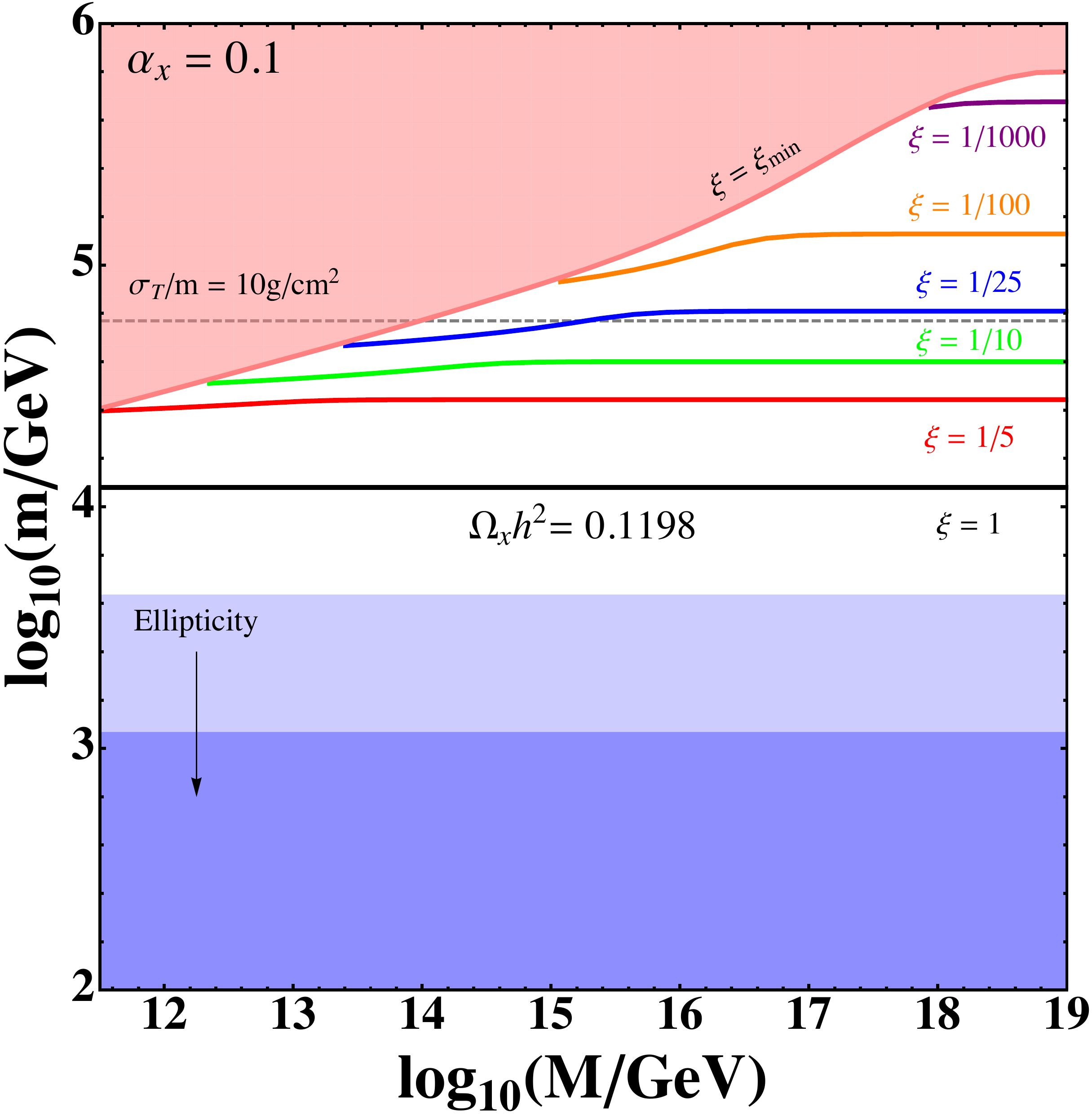}
        \hspace{0.5cm}
        \includegraphics[width=0.45\textwidth]{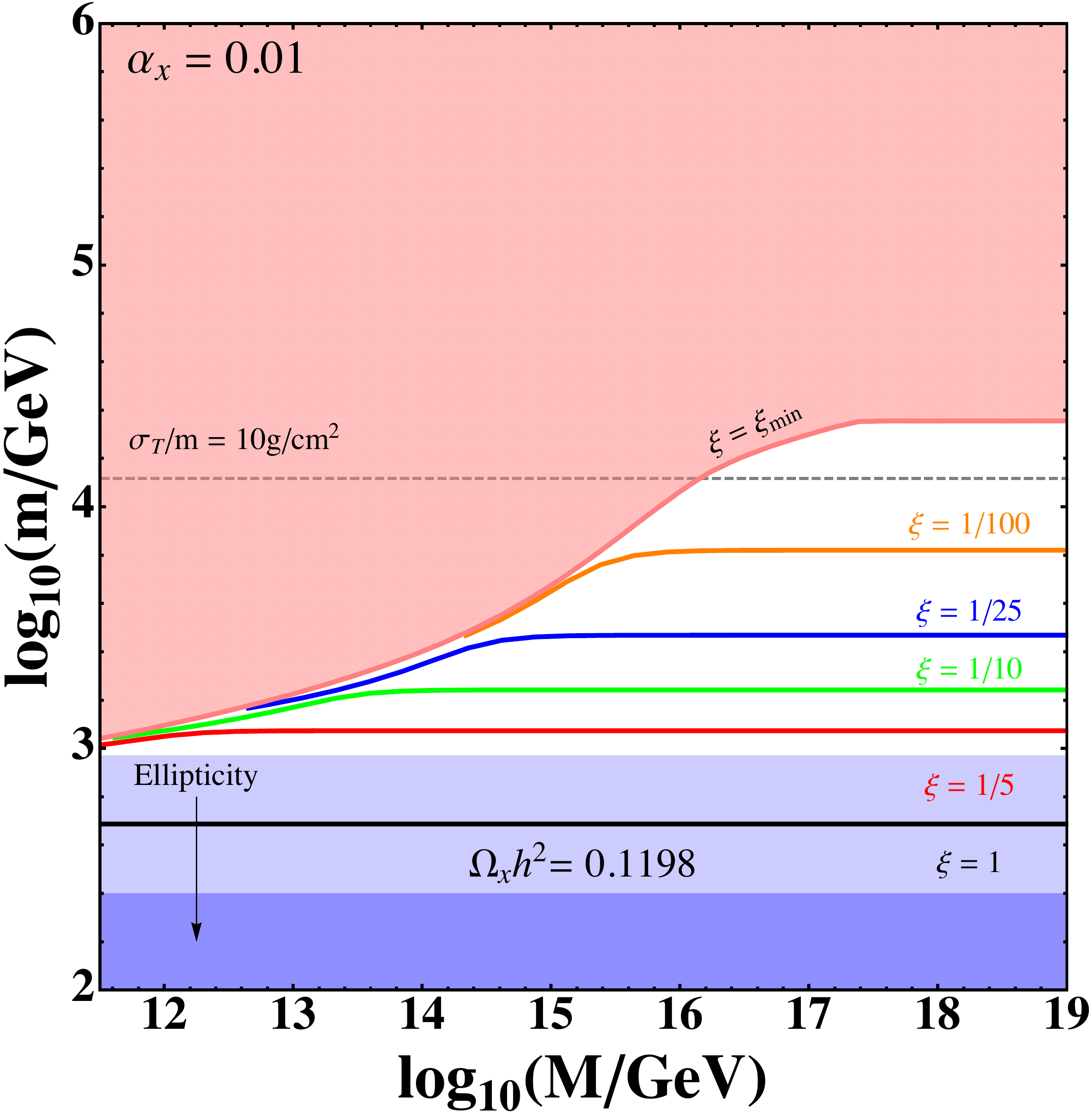}
	\caption{
Values of $\mpsi$ that give the correct relic density of $\psi$ dark matter
as a function of $M$ for $\alpha_x = 0.1$~(left) and $0.01$~(right)
for various fixed values of $\xi$.  Each solid line corresponds to
the correct $\psi$ relic density for the corresponding value of $\xi$.
The red shaded upper region is excluded due to overproduction of $\psi$ relic
density for any consistent value of $\xi$.  The lower blue shaded regions 
indicate exclusions from the effects of $\psi$ dark matter self-interactions
from the observed ellipticity of galactic halos, with the dark blue indicating
a conservative exclusion and the light blue showing a more aggressive one.
The dotted line indicates a DM self-scattering transfer cross-section per mass 
in dwarf halos of $\sigma_T/m_\psi = 10\,\mathrm{cm}^2/\mathrm{g}$.
}
	\label{fig:reld}
\end{figure}

  Ultimately, we are interested in the parameter space where $\psi$ 
can make up all the dark matter.  In Fig.~\ref{fig:reld} we show
the values of $\mpsi$ for which this occurs as a function of $M$
for $\alpha_x = 0.1$~(left) and $0.01$~(right) for various values 
of $\xi$~(solid lines).  
The lines in these plots are cut off at smaller $M$ when 
$\xi$ falls below $\xi_{min}$.
As expected from the annihilation cross section, larger values
of $\alpha_x$ coincide with larger dark matter masses.
In the right part of both panels the allowed DM mass $\mpsi$
reaches a value that is independent of $M$ for fixed $\xi$.
This region corresponds to late transfer being negligible for 
the freeze-out process, with the relic density scaling approximately
as $\xi^{-1}\alpha_x^2/\mpsi^2$.  Going to smaller $M$, transfer eventually
becomes important and the relic density increases.  Correspondingly,
the mass $\mpsi$ that produces the correct relic density decreases.
As $M$ decreases further, the lines for different $\xi$ values
in Fig.~\ref{fig:reld} come together.  This can be understood
from Eqs.~\eqref{eq:xxdec} and \eqref{eq:omx2}, where the direct dependence
on $\xi$ is seen to cancel for cross sections 
$\langle\sigma_{ann} v\rangle = \sigma_0x^{-n}$ with $n\to 0$, as we have
here (up to the Sommerfeld and bound state enhancements).
The upper shaded region in both panels is excluded because the
resulting relic density of $\psi$ is always greater than the 
observed DM density for any consistent value of $\xi$.
Going from $\alpha_x=0.1$ to $0.01$, lower $\psi$ masses
are needed to produce the correct relic density.
Also shown in this figure are bounds from DM self-interactions 
to be discussed below.

\section{Dark Matter Self-Interactions\label{sec:dmself}}

  Dark matter in our theory is charged under an unbroken
$U(1)_x$ gauge force implying long-range self interactions among DM particles
that can modify their behavior in collapsed systems.  Such interactions 
have been suggested as a way to resolve several apparent discrepancies
between simulations of DM structure formation and 
observations~\cite{Spergel:1999mh,Tulin:2017ara}.
However, these interactions are also constrained to not be so large
as to overly disrupt cosmic structures~\cite{Ackerman:mha,Tulin:2013teo}.

  An upper bound on DM self-interactions can be derived from
the observed ellipticity of galactic halos
such as NGC720~\cite{Buote:2002wd,Humphrey:2010hd}. 
For charged DM coupled to an unbroken $U(1)$, 
Refs.~\cite{Feng:2008mu,Agrawal:2016quu}
derived limits on the gauge coupling of the form
\beq
\alpha_x \lesssim \left\{0.35,\,2.5\right\}\times 10^{-6}\;
\lrf{\mpsi}{\gev}^{3/2} \ ,
\label{eq:alfxmax}
\eeq
where the two numbers in brackets correspond to the analyses of 
Refs.~\cite{Feng:2008mu} and \cite{Agrawal:2016quu}, respectively.  
While the limit derived in Ref.~\cite{Feng:2008mu} is considerably stronger, 
Ref.~\cite{Agrawal:2016quu} (and Ref.~\cite{Kahlhoefer:2013dca}) 
argue for a weaker one based on the application of the ellipticity
constraint only at larger galactic radii and a number of smaller factors.
We show both upper bounds on $\alpha_x$ in Figs.~(\ref{fig:reld}).
These favor smaller temperature ratios $\xi$ and larger DM masses $\mpsi$,
well above the weak scale.

  The limits on $\alpha_x$ from the ellipticity of NGC720
correspond to an effective transfer cross section per mass below about
$\sigma_T/\mpsi \lesssim 1\,\text{cm}^2/\text{g}$ in this system with
a velocity dispersion on the order of $v \simeq 300\,\text{km/s}$.
Dark matter self-interactions in this regime are described by 
a Rutherford-like transfer cross section~\cite{Feng:2008mu,Kahlhoefer:2013dca,Agrawal:2016quu}:
\beq
\sigma_T ~\simeq~ \frac{8\pi\,\alpha_x^2}{\mpsi^2}\,\frac{1}{v^4}\,\ln\Lambda \ ,
\eeq
where $\ln\Lambda \sim 45\!-\!75$ is a collinear enhancement factor cut off
by the typical interparticle spacing in the system~\cite{Agrawal:2016quu}.  
Since this cross section has a very strong velocity dependence, 
the DM self-interactions in systems with lower velocity 
dispersions such as dwarf halos can be much stronger.  Using typical
velocities and densities for dwarf halos,
this translates into
\beq
\sigma_T/\mpsi ~\simeq~ 18\,\text{cm}^2/\text{g}\,
\lrf{\alpha_x}{0.1}^2
\lrf{5\times 10^4\,\gev}{\mpsi}^3\lrf{10\,\text{km/s}}{v}^4\lrf{\ln\Lambda}{50}
\label{eq:dwhalo}
\eeq
Interaction cross sections of this size are expected to lead to the formation 
of cores in dwarf halos, with Refs.~\cite{Vogelsberger:2012ku,Zavala:2012us} 
suggesting a better agreement between simulations
and data for $\sigma_T/\mpsi \sim 10\,\text{cm}^2/\text{g}$.
On the other hand, it is not clear what the upper bound on
$\sigma_T/\mpsi$ is from these systems, 
with the simulations of Ref.~\cite{Elbert:2014bma}
finding reasonable behavior for  $\sigma_T/\mpsi = 50\,\text{cm}^2/\text{g}$
(the largest value studied) and Ref.~\cite{Agrawal:2016quu} arguing that
much larger values can work as well.  
Indeed, the results of Ref.~\cite{Elbert:2014bma} appear to be consistent 
with the approximate duality between $\sigma_T/\mpsi$
and $\mpsi/\sigma_T$ about Knudsen number close to unity suggested
in Ref.~\cite{Agrawal:2016quu} based on the analyses of
Refs.~\cite{Ahn:2002vx,Ahn:2004xt}.  For reference,
we also show dashed contours indicating 
$\sigma_T/\mpsi = 10\,\text{cm}^2/\text{g}$
in Figs.~\ref{fig:reld}.

\section{Conclusions\label{sec:conc}}

The standard expectation for non-renormalizable operators in the early universe
is that their effects are greatest at high temperatures and that they
decouple at lower temperatures.  For this reason, DM creation from
SM collisions connecting to a secluded dark sector through a non-renormalizable
operator is referred to as UV freeze-in~\cite{Hall:2009bx,Elahi:2014fsa}.
In this work we showed that such operators can also contribute importantly
at lower temperatures when combined with freeze-out in a dark sector.
  
  To illustrate the effect, we studied a simple dark sector consisting
of a massive Dirac fermion $\psi$ DM candidate 
and a massless Abelian dark vector $X^{\mu}$,
with the only connection to the SM through the dimension-five 
fermionic Higgs portal operator of Eq.~\eqref{eq:fhp}.  At the end of
reheating, the dark sector can be populated by transfer reactions 
$\text{SM}+\text{SM} \to \psi+\bar{\psi}$ mediated by the non-renormalizabel 
portal operator to a density below the value it would have in full equilibrium 
with the SM.  As the universe cools further, the population of dark fermions 
can equilibrate with the dark vectors at temperature $T_x$ below the visible
SM temperature $T$ provided the dark gauge coupling and the initial
fermion density are large enough.  Freeze-out occurs in the dark sector
when $T_x$ falls below the fermion mass $\mpsi$.  For a broad range of 
parameters in this theory, the relic density of $\psi$ fermions can receive
a significant additional enhancement from late transfer reactions
through the non-renormalizable portal operator during the course of the 
freeze-out process for $T$ down and below the fermion mass.  
The UV connector operator of Eq.~\eqref{eq:fhp} is therefore seen to
play an important role in the IR.

  The simple dark sector theory we considered also has interesting implications 
for DM self-interactions, which are motivated by a number of puzzles in cosmic
structure~\cite{Spergel:1999mh,Tulin:2017ara}.  Such interactions were
investigated for this theory in Refs.~\cite{Feng:2008mu,Kahlhoefer:2013dca,Agrawal:2016quu} and suggest that to be viable larger DM masses and smaller 
temperature ratios $\xi = T_x/T$ are required to avoid bounds from the observed
ellipticity of NG720.  These bounds, and the dependence of the self-interaction
cross section on the DM velocity, could potentially be softened by extending 
the theory to include a small mass for the dark vector~\cite{Tulin:2013teo}.  
The calculations presented in this work can be carried over to such a 
massive vector scenario provided its mass is much smaller than the decoupling
temperature of the dark fermion so that it provides a relativistic thermal
bath during this process.  Furthermore, the vector mass would also have
to be small enough to avoid too much vector 
boson DM~\cite{Ma:2017ucp,Duerr:2018mbd}.  

  While this work focused on a specific dark sector theory 
and non-renormalizable connector operator, a similar IR contribution
from a non-renormalizable connector to the density of dark-sector DM is
expected to occur as well for other dark sectors or connector operators.
For the effect to arise, the DM candidate in the dark sector must undergo 
significant annihilation to allow the power-suppressed transfer reactions 
(relative to reheating) of the connector operator to catch up.  
Other non-renormalizable connector operators can also lead to late IR transfer
contributions to the DM relic density, although initial estimates
suggest that the effect becomes less important as the operator dimension
increases.  Late-time transfer of a symmetric density could also be 
relevant in scenarios of secluded asymmetric DM.

  Dark matter arising from a dark sector that is colder than the SM
in the early universe has been investigated in a wide range 
of scenarios of new physics~\cite{Feng:2008mu,Das:2010ts,Elahi:2014fsa,Agrawal:2016quu,Krnjaic:2017tio,Faraggi:2000pv,CyrRacine:2012fz,Boddy:2014yra,Boddy:2014qxa,Berlin:2016vnh,Berlin:2016gtr,Soni:2016gzf,Forestell:2016qhc,Halverson:2016nfq,Acharya:2017szw,Forestell:2017wov,Krnjaic:2017tio,Bernal:2018qlk}.  
In some of these works, the dark temperature $T_x$ is taken as an input 
to the calculation of the DM relic density without reference to how 
the dark sector was populated initially.  Our results show that such
an assumption is not always justified, and the nature of the connector
operators that mediate transfer from the SM to the dark sector can play
an important role in determining the relic density of DM.

\section*{Acknowledgements}

We thank Dave McKeen, Nirmal Raj, Kris Sigurdson, Sean Tulin, 
Graham White, and Yue Zhang for helpful discussions.
DEM acknowledges the Aspen Center for Physics, which is funded by National 
Science Foundation grant PHY-1607611, for their hospitality
while this work was being completed.
This work is supported by the Natural Sciences
and Engineering Research Council of Canada~(NSERC), with DEM 
supported in part by Discovery Grants and LF by a CGS~D scholarship.
TRIUMF receives federal funding via a contribution agreement 
with the National Research Council of Canada.

\appendix

\section{Calculation of Transfer Rates\label{sec:appa}}

  In this appendix we calculate the effective transfer rates
of number and energy density from the visible sector to the 
dark sector through the operator of Eq.~\eqref{eq:fhp}.
The squared matrix element for $\psi+\bar{\psi}\to H+H^{\dagger}$
derived from this interaction and summed over both initial
and final degrees of freedom is
\beq
\widetilde{|\mathcal{M}|^2} = \frac{4}{M^2}(s-4\mpsi^2) \ ,
\label{eq:matel}
\eeq
with $s=(p_1+p_2)^2$.  Note that we assume implicitly that the
Higgs is in the electroweak unbroken phase and can be treated
as a massless $SU(2)_L$ scalar doublet.

\subsection{Number Transfer}

  The relevant number transfer term via
$\psi(1)+\bpsi(2)\to H(3)+H^{\dagger}(4)$ is
\beq 
\mathcal{T}(T)
&~\equiv~& 
\langle\sigma_{tr} v(T)\rangle(\npsi^2-\npsieq^2(T))
\label{eq:ctrn0}\\
&~\equiv~&
\int\!d\Pi_1\!\int\!d\Pi_2\!\int\!d\Pi_3\!\int\!d\Pi_4\;
(2\pi)^4\delta^{(4)}(p_i)
\widetilde{|\mathcal{M}|^2}(f_1f_2-f_3f_4) \ ,
\nnmb
\eeq
where $d\Pi_i = d^3p_i/2E_i(2\pi)^3$.  To make the calculation tractable,
we approximate the distribution functions by the Maxwell-Boltzmann
form $f_i = \zeta_ie^{-E_i/T}$, where $\zeta_i$ is the rescaling needed 
to get the correct number densities relative to equilibrium at temperature $T$.
We expect that the Maxwell-Boltzmann approximation used here is correct
up to factors very close to unity.  

For $\npsi \ll \npsieq(T)$ we have $f_1 = f_2 \simeq 0$, while Higgs fields 
in full thermodynamic equilibrium with the SM (in the electroweak unbroken phase)
imply $f_3=f_4 = 1$.  The transfer term then reduces to
\beq
\mathcal{T}(T) = \int\!d\Pi_1\!\int\!d\Pi_2\;
\left(4g_\psi^2E_1E_2\sigma_{tr}v\right)e^{-(E_1+E_2)/T} \ ,
\eeq
with $g_\psi=2$ being the number of fermion spin states.
Note that the combination in brackets is Lorentz invariant and can depend only
on the variable $s$.  It is given by
\beq
\left(4g_\psi^2E_1E_2\sigma_{tr}v\right)
&=& \int\!d\Pi_3\!\int\!d\Pi_4\;(2\pi)^4\delta^{(4)}(p_i)
\widetilde{|\mathcal{M}|^2}
\\
&=& \frac{1}{8\pi}
\left(\frac{1}{4\pi}\int\!d\Omega\,\widetilde{|\mathcal{M}|^2}\right)_{CM}
\nnmb\\
&=& \frac{1}{2\pi}\frac{1}{M^2}(s-4\mpsi^2) \ .
\nnmb
\eeq
To integrate this over the initial states, we follow 
Refs.~\cite{Gondolo:1990dk,Edsjo:1997bg}
and use the fact that the integrand depends only on $s$ and $E_+ = (E_1+E_2)$
to write
\beq
\int\!d\Pi_1\int\!d\Pi_2 &=& \frac{1}{4(2\pi)^4}\int_{4\mpsi^2}^\infty\!ds
\int_{\sqrt{s}}^{\infty}dE_+\,\sqrt{1-4\mpsi^2/s}\sqrt{E_+^2-s} \ . 
\eeq
Since the only $E_+$ dependence of the integrand is in the Boltzmann exponential,
integrating using a Bessel function identity\footnote{
$K_{\nu}(z) = \frac{\sqrt{\pi}z^\nu}{2^\nu\Gamma(\nu+1/2)}
\int_1^{\infty}\!dt\,(t^2-1)^{\nu-1/2}e^{-zt}$.}
gives
\beq
\mathcal{T}(T) &=& 
\frac{1}{4(2\pi)^4}\int_{4\mpsi^2}^{\infty}\!ds\,
(4g_1g_2E_1E_2\sigma_{tr}v)\,\sqrt{1-4\mpsi^2/s}
\label{eq:ctrn1}
\\
&=& 
\frac{1}{2(2\pi)^5}\frac{T^6}{M^2}\mathcal{F}(x) \ ,
\nnmb
\eeq
where $x=\mpsi/T$ and
\beq
\mathcal{F}(x) &=& \int_{2x}^{\infty}\!du\,u\,(u^2-4x^2)^{3/2}\,K_1(u)\\
&\simeq& \left\{
\begin{array}{lcl}
16&~~;~~~&x\ll 1
\nnmb\\
&&\label{eq:ntrans}\\
6\pi\,x^2e^{-2x}&~~;~~~&x\gg 1
\end{array}\right.
\nnmb
\eeq

\subsection{Energy Transfer}

  We are also interested in the net rate of energy transfer between
the visible and dark sectors.  The relevant energy collision term
for $\psi+\bpsi\to H+H^\dagger$ is identical to Eq.~\eqref{eq:ctrn0}
but with an additional factor of $\Delta E = (E_1+E_2) = E_+$
in the integrand.  The result is\footnote{
$\int_{1}^\infty\!dt\,t\sqrt{t^2-1}\,e^{-zt} = 
-\frac{d}{dz}[K_1(z)/z] = K_2(z)/z$.}
\beq
\mathcal{U}(T) &\equiv& \langle \Delta E\cdot\sigma_{tr}v(T)\rangle
(\npsi^2-\npsieq^2(T))
\nnmb\\ 
&~\equiv~&
\frac{T}{4(2\pi)^4}\int_{4\mpsi^2}^{\infty}\!ds\,
(4g_1g_2E_1E_2\sigma_{tr}v)
\sqrt{s-4\mpsi^2}\;\sqrt{s}K_2(\sqrt{s}/T)
\label{eq:ctre1}\\
&=& \frac{1}{2(2\pi)^5}\frac{T^7}{M^2}\mathcal{G}(x)
\nnmb
\eeq
with $x=\mpsi/T$ and
\beq
\mathcal{G}(x) &=& \int_{2x}^{\infty}\!du\,u^2\,(u^2-4x^2)^{3/2}\,K_2(u)
\nnmb\\
&&\label{eq:etrans}\\
&\simeq& \left\{
\begin{array}{lcl}
96&~~;~~~&x\ll 1
\\
12\pi\,x^3e^{-2x}&~~;~~~&x\gg 1
\end{array}\right.
\nnmb
\eeq


\bibliography{ref_inspire}

\end{document}